\documentclass[9pt,twocolumn,twoside]{osajnl}
\usepackage{comment}
\usepackage{soul}
\journal{ol} 
\setboolean{shortarticle}{true}

\title{Does the structure of light influence the speckle size?}
\author[1,2]{Xiao-Bo Hu}  
\author[2]{Meng-Xuan Dong}
\author[2]{Zhi-Han Zhu}
\author[2]{Wei Gao}
\author[2,*]{Carmelo Rosales-Guzm\'an}
\affil[1]{The Higher Educational Key Laboratory for Measuring \& Control Technology and Instrumentations of Heilongjiang Province, Harbin University of Science \& Technology, Harbin 150080, China}
\affil[2]{Wang Da-Heng Collaborative Innovation Center, Heilongjiang Provincial Key Laboratory of Quantum manipulation \& Control, Harbin University of Science and Technology, Harbin 150080, China}

\affil[*]{Corresponding author:carmelorosalesg@hrbust.edu.cn}

\begin{abstract}
It is well known that when a laser is reflected from a rough surface or transmitted through a diffusive medium, a speckle pattern will be formed at a given observation plane. Speckle is commonly produced by laser beams with a homogeneous intensity, for which, well-known relations have been derived, relating the speckle size to the area of illumination. Here we investigate the speckle generated by higher-order Laguerre-Gaussian (LG) modes, characterized by a non-uniform intensity distribution of concentric rings.We show that the ring-structure of the LG modes does not play any role in the speckle size, which happens to be the same as that obtained for a homogeneous intensity distribution. This allow us to provide with a simple expression that relates the speckle size to the spot size of the LG modes. Our findings will be of great relevance in many speckle-based applications. 
\end{abstract}

\setboolean{displaycopyright}{true}

\begin{document}

\maketitle
As almost as early as the invention of the laser, an unexpected phenomena was observed when a rough surface was illuminated by a laser, the formation of a grainy structure of bright spots of light and darkness: speckle. Interestingly, this phenomenon had long been known by Newton but it was only after the invention of highly coherent light sources that its observation became more common. Incidentally, speckle can also be observed when a laser is transmitted through a medium with random refractive index fluctuations \cite{Piederriere2005}. Even though in its origins laser speckle was regarded as an undesired noise that should be removed, specially for digital holography \cite{Bianco2018}, it has turned into a powerful optical tool that has found applications in fields, such as, optical metrology, imaging or medicine, to mention a few \cite{SanJuan2008,SanJuan2013,Francon1979,Cheng2007,Fercher1981,Heeman2019,Zakharov2009,Dunn2003}. The speckle phenomena can be easily understood in terms of the microscopic structure of the surface if we invoke the Huygens principle \cite{Goldfischer1965,Progresinoptics14}. In essence, if the roughness of the illuminated surface is in the order of the wavelength of illumination, its height variations will induce random phase differences in the scattered light. Hence, the optical wave at a given distance may be considered as the superposition of a large number of coherent wavelets, each arising from the various microscopic elements on the surface, whose phases are in essence random. In this way, the intensity of the scattered field will be constituted by the locus of bright spots, of constructive interference, interlaced with dark spots, of destructive interference. 

In recent time, with the increasing development of novel tools to engineer the structure of light in an almost unlimited fashion \cite{rosales2017shape}, triggered by the many applications it spans \cite{Roadmap,TwPh,Hell1994,RosalesGuzman2013,Trichili2016,Willner2015}, the need for understanding the interaction of structured light with scattering mediums has become obvious. For example, a proper understanding of structured light through atmospheric turbulence is of great relevance for free-space optical communications \cite{Cox2016,Malik2012,Cox2018}. An important parameter of speckle is its size, and up to now, very little has been said in relation to this and the structure of the illuminating source. Along this line, pioneering studies have considered optical vortices, which, features a ring-like intensity profile and an azimuthal phase distribution of the form $\exp{i\ell\phi}$. Here $\ell \in \mathbb{Z}$, known as the topological charge, accounts for number of times the phase wraps around the optical axis. Incidentally, to these optical vortices is associated an amount $\ell\hbar$ of Orbital Angular Momentum (OAM) per photon \cite{Allen92}. In these early studies, it was reported a decrease in the speckle size associated to an increase on the topological charge. Here, it was already suggested that the area of the ring-like intensity illuminating the scattering surface, which increases with the topological charge, was responsible for a decrease in the speckle size \cite{Reddy2014}. Hence, the authors proposed an expression that relates the ring-like area of the intensity to the size of the speckle, which in essence suggest that the structure of the illuminating area plays a  role in determining the size of the speckle. Further studies used optical vortices of constant size, so-called perfect vortex beams, to corroborate that the decrease of speckle size is complete unrelated to the topological charge \cite{Reddy2016}. 

Here, we consider a more general class of structured beams, the set of Laguerre-Gaussian ($LG_p^\ell$) modes, characterized by a complex intensity profile formed by $p-1 (p \in {\mathbb N})$ concentric rings of varying intensity and topological charge $\ell$, to establish a general relation between the speckle size and the $LG_p^\ell$ modes. First, we show that as $p$ and $\ell$ increase, the speckle size decreases, even for the case $\ell=0$, which reinforces the idea that the OAM does not influence the speckle size. More importantly, our study evinces that the decrease in speckle size is uniquely determined by an enlargement of the illuminated area rather than to the complexity of the beam's structure. This allow us to provide with a specific expression that relates the speckle size to the  modal indices $p$ and $\ell$. To corroborate that the ring-like structure of the modes does not play any role in determining the speckle size, we compare our results with homogeneous-intensity apertures, finding a perfect match between our theoretical expressions and both illuminating structures. 

To start with, let us remind that set of $LG_p^\ell$ modes are given, in the cylindrical coordinates ($\rho, \varphi$), by \cite{RosalesGuman2017a}, 
\begin{align}
\nonumber
LG_p^\ell (\rho,\varphi,z)=& \sqrt{\frac{2p!}{\pi(|\ell|+p)!\omega^2(z)}}\left[\frac{\sqrt{2}\rho}{\omega(z)}\right]^{|\ell|}L_p^{|\ell|}\left[\frac{2\rho^2}{\omega^2(z)}\right]\\\nonumber
&\exp[i(2p+|\ell|+1)\zeta(z)]\exp\left[-\frac{\rho^2}{\omega^2(z)}\right]\\
&\exp\left[\frac{-ik\rho^2}{2R(z)}\right] \exp[-i\ell\varphi]\exp[-ikz],
\label{LG}
\end{align}
where, $L_p^\ell(x)$ are the generalized Laguerre Polynomials. The radius of curvature $R(z)$, the Gaussian beam radius $\omega(z)$ and the Gouy phase $\zeta(z)$, are defined as,
\begin{align}
\nonumber
&R(z)=z\left[1+\left(\frac{z_R}{z}\right)^2\right],\hspace{5mm}\omega(z)=\omega_0\sqrt{1+\left(\frac{z}{z_R}\right)^2},\\
&\textrm {and} \hspace{5mm} \zeta(z)=\arctan\left(\frac{z}{z_R}\right),
\label{LGparam}
\end{align}
respectively. Moreover, $\omega_0$ is the Gaussian beam waist at $z=0$ and $z_R=\pi\omega_0^2/\lambda$ is the Rayleigh range of the Gaussian mode.

\begin{figure}[t]
    \centering
    \includegraphics[width=.49\textwidth]{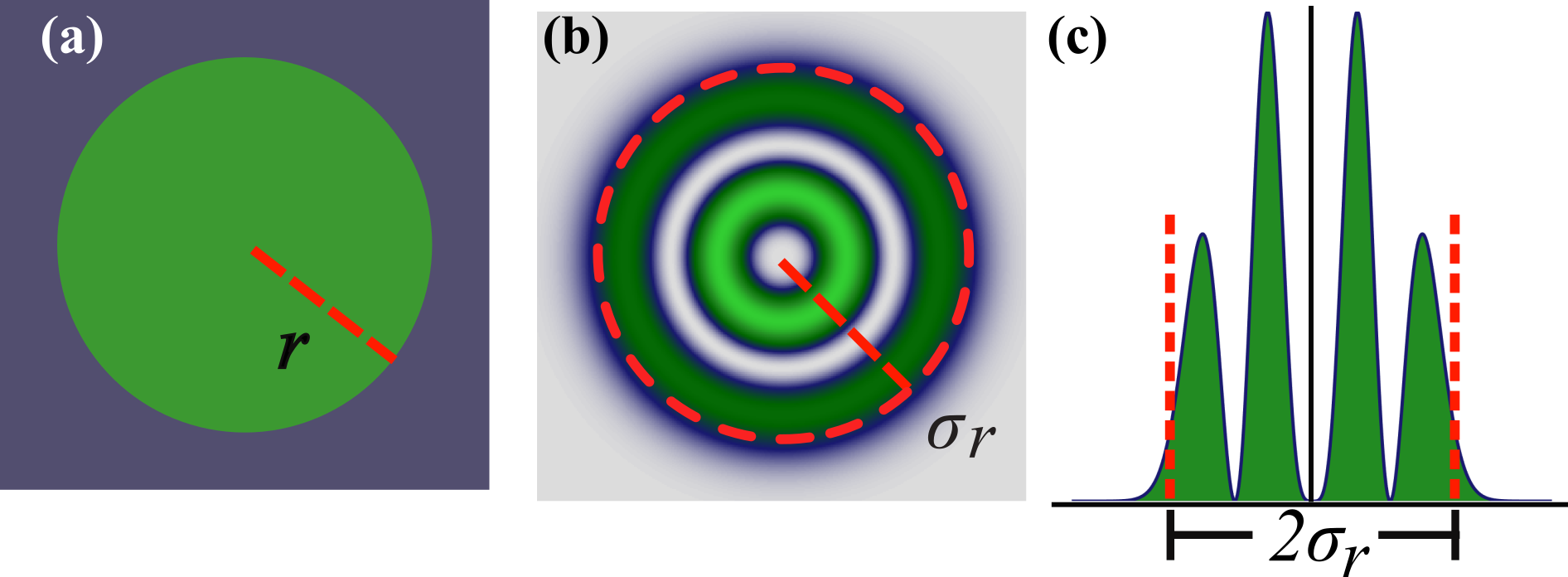}
    \caption{ (a) Circular aperture of radius $r$. (b) Intensity profile of an $LG_p^\ell$ mode of spot size radius $\sigma_r$, defined as the maximum radial distance containing all the intensity maximum. (c) Transverse section of the intensity profile showed in (c).}
    \label{LGdisc}
\end{figure}

To find the speckle size when a rough surface is illuminated by a laser beam, as the one described by Eq. \ref{LG}, we first have to emphasize that speckles do not have a  well defined size, all we can do is to provide with the mean size. A common method to compute this is through the normalized autocorrelation function of the speckle intensity $I(u,v)$ at a given observation plane $z_{obs}$, that is,
\begin{equation}
C(\Delta u, \Delta v)=\big\langle I(u_1,v_1)I(u_2,v_2) \big\rangle, 
\label{AutocorrDef}
\end{equation}
where, $\Delta u=u_2-u_1$, and $\Delta v=v_2-v_1$. Further, $I(u_i,v_i)$, $i=1,2$, is the intensity at two points in the observation plane and $\big\langle\cdot \big\rangle$ represents the spatial average over a large number of speckles. The mean speckle size is then defined as the distance $\Delta r$ at which both intensities become uncorrelated to one another. A mathematical expression for  $C(\Delta u, \Delta v)$, first derived by Goodman \cite{Goodman1975}, is given by,
\begin{align}
\nonumber
    &C(\Delta u, \Delta v)=\\
   &\big\langle I\big\rangle^2 \left[1+\left|\frac{\iint\limits_{-\infty}^{+\infty} |P(x,y)|^2 \exp \left[\frac{i2\pi}{\lambda z_{obs}}(x\Delta u+y\Delta v)\right] \mathrm{d}x \mathrm{d}y }{\iint\limits_{-\infty}^{+\infty} |P(x,y)|^2 \mathrm{d}x \mathrm{d}y }\right|^2 \right]
    \label{Autocorr}
\end{align}
where $P(x,y)$ represents the amplitude of the field incident on the scattering surface. For the specific case of a unit amplitude plane wave illuminating a circular region of radius $r$ given by,
\begin{equation}
|P(x,y)|^2=\mathrm{circ}\left(\frac{\sqrt{x^2+y^2}}{r}\right),
\label{IntDisc}
\end{equation}
Eq. \ref{Autocorr} will reduce to \cite{MatlabTutorial},
\begin{equation}
    C(\Delta u, \Delta v)=\big\langle I\big\rangle^2 \left[1+\frac{r^3}{\lambda z_{obs}}\left|\frac{J_1\left(\frac{2\pi r}{\lambda z}\sqrt{\Delta u^2+\Delta v^2}\right)}{\sqrt{\Delta u^2+\Delta v^2}}\right|^2 \right],
    \label{AutocorrLG}
\end{equation}
where  $J_1$ is the Bessel function of the first kind and order 1. The mean speckle radius of a speckle can then be taken as the value $\sqrt{\Delta u^2+\Delta v^2}=\Delta s$ for which $J_1(x)$ first becomes zero, which happens when the argument is equal to $1.22\pi$. Hence, the mean speckle diameter will be, 
\begin{equation}
\Delta s=\frac{1.22\lambda z_{obs}}{2r}, 
\label{SpSize}
\end{equation}
This is a very famous relation that shows the average speckle size increases linearly with the distance from the scattering surface to the observation plane and decreases as the illuminated area increases. 
To measure the speckle size generated by the $LG_p^\ell$ modes, we should, in principle, compute Eq. \ref{Autocorr} with the amplitude distribution $P(x,y)$ given by Eq. \ref{LG}. Nevertheless, as we will demonstrate, the speckle size can be measured very accurately by approximating the complex intensity distribution by a circular aperture of uniform intensity (Eq. \ref{IntDisc}). For this, we have to first compute the total area of the $LG_p^\ell$ modes illuminating the rough surface taking into account its increase with $|\ell|$ and $p$ (see Fig. \ref{LGdisc} (b)). This can be done through the generalized spot size, which is defined as the maximum area to where the beam's intensity still has a significant value (and \ref{LGdisc} (c)). Following this definition, a mathematical expression for the radius $\sigma$ of the spot size can be derived by using the standard deviation as,
\begin{eqnarray}
    \sigma^2(z)=\frac{2\int_0^{2\pi}\int_{0}^\infty \rho^2 I(\rho,\varphi,z) \textrm{d}\rho \textrm{d}\varphi}{\int_0^{2\pi}\int_{0}^\infty I(\rho,\varphi,z) \textrm{d}\rho\textrm{d}\varphi},
    \label{LGsigma}
\end{eqnarray}
where, $I(\rho,\varphi,z)$ is the intensity of the $LG_p^\ell((\rho,\varphi,z)$ modes (Eq. \ref{LG}). After computing both integrals, Eq. \ref{LGsigma} reduces to \cite{Phillips1983},
\begin{eqnarray}
    \sigma(z)=\omega(z)(2p+|\ell|+1)^{1/2},
    \label{LGwaist}
\end{eqnarray}
Combination of Eq. \ref{SpSize} and Eq. \ref{LGsigma} provides then with the diameter of the mean speckle size for $LG_p^\ell$ modes, namely,
\begin{equation}
\Delta s=\frac{1.22\lambda z_f}{2 \omega(z)(2p+|\ell|+1)^{1/2}},
\label{LGss}
\end{equation}
which is a very simple expression that relates the speckle size to the spot size of the $LG_p^\ell$ modes.

\begin{figure}[tb]
\centering
\includegraphics[width=.5\textwidth]{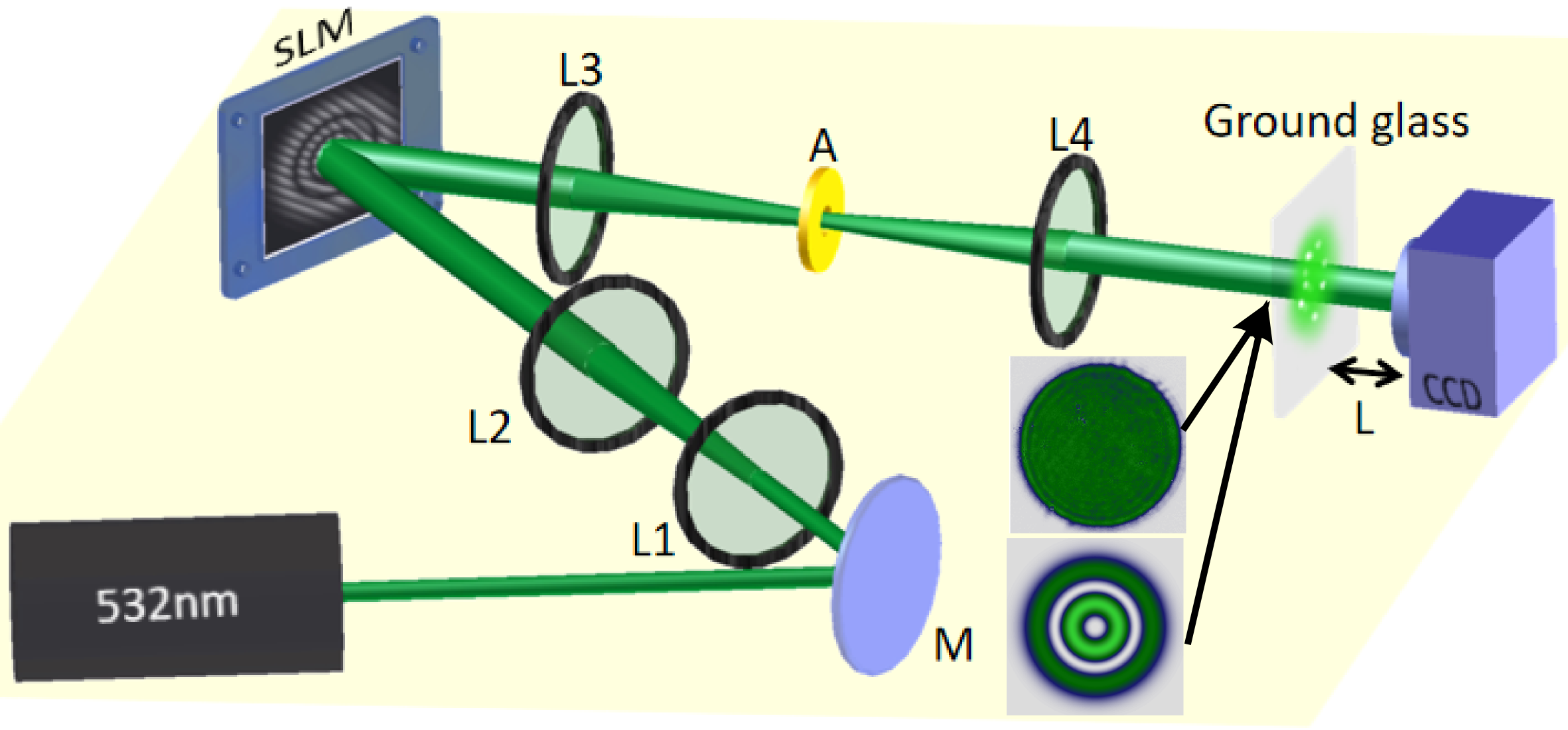}
\caption{Schematic representation of the experimental setup implemented to measure the mean speckle size. M: Mirror, SLM: Spatial Light Modulator, L1-L4: Lens, P: Pinhole, CCD: Charge-Coupled Device camera. The insets illustrate the intensity profile of the beams illuminating the ground glass at the plane $z=0$.}
\label{exp}
\end{figure}

\begin{figure}[tb]
\centering
\includegraphics[width=.46\textwidth]{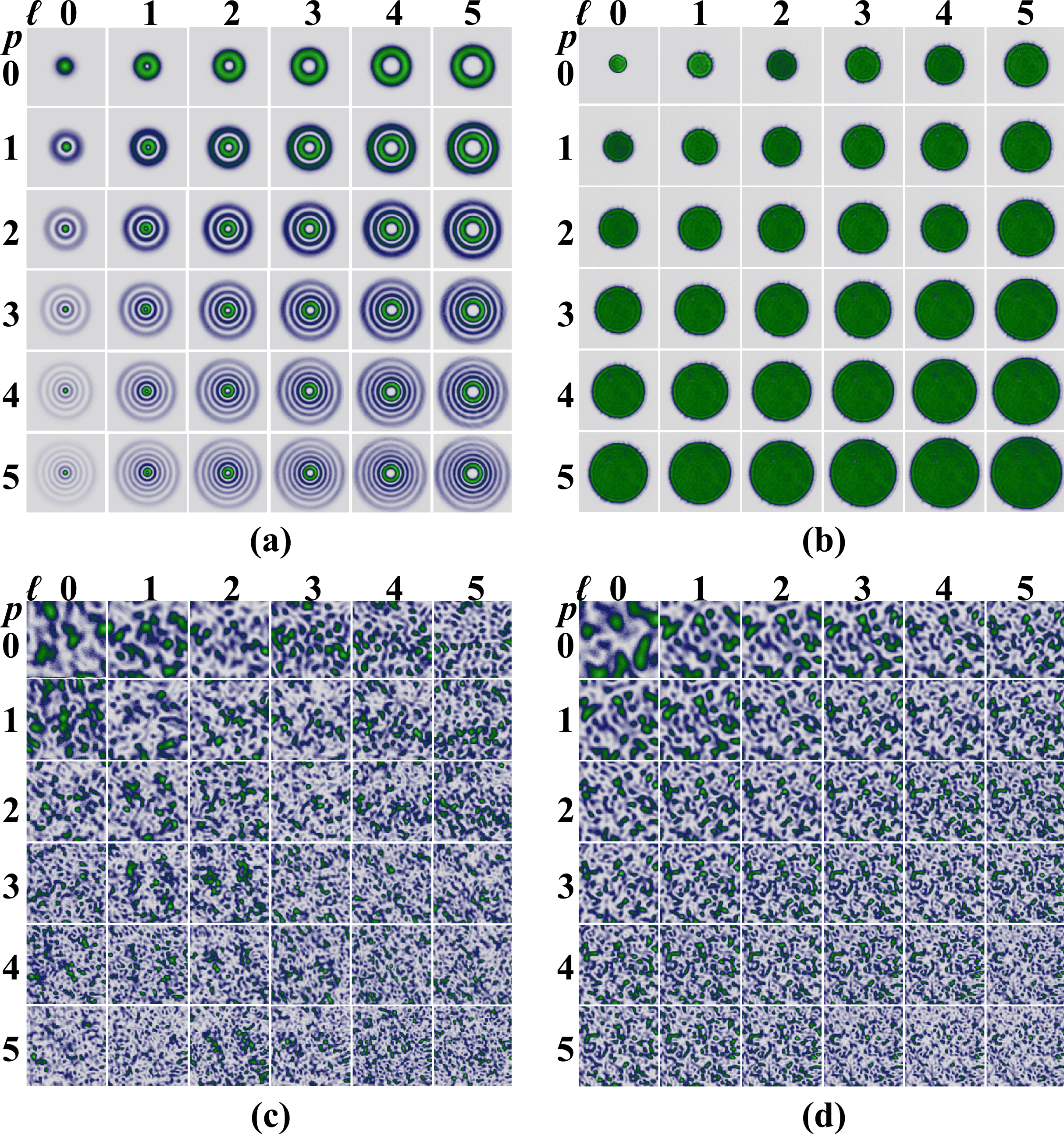}
\caption{(a) Intensity profile for a set of apertures with varying radius $r$. (b) Intensity profile of the subset of $LG_p^\ell$ modes given by combinations of $p\in[0,5]$ and $\ell\in[0,5]$. (c) Speckle intensity produced by the intensity distribution of the apertures in (a). (d) Speckle intensity produced by the $LG_p^\ell$ showed in (b).}
\label{LGIntensity}
\end{figure}

In order to prove the validity of Eq. \ref{LGss}, we performed an experimental corroboration using the experimental setup depicted in Fig. \ref{exp}. A continuous wave (CW) laser beam at $\lambda=532$ nm was expanded, using lenses L1 ($f=25mm$) and L2 ($f=150mm$), to approximate a flat wave front and directed onto a Spatial Light Modulator (SLM, Holoeye Pluto $1920x1080$, $8\mu$m pixel size) located at the plaze $z=0$. The $LG_p^\ell$ modes were encoded on the SLM using a complex amplitude modulation approach \cite{RosalesGuman2017a,rosales2017shape}. The beam waist for all $LG_p^\ell$ modes was chosed as $\omega_0=300\mu$m. Afterwards, the modes encoded on the SLM were expanded two times and imaged onto the plane of the scattering surface, a ground glass plate, using a 4f imaging system composed of lenses L3 ($f=100mm$) and L4 ($f=200mm$). The purpose of the image system is to relay the plane $z=0$ to the plane where the ground glass in placed, in other words, it allows us to use $\omega(z)$ as $\omega_0$. In addition, a spatial pinhole placed between L3 and L4 enabled the removal of higher diffraction orders. 
Figure \ref{LGIntensity}(a) show the intensity distribution of a subset of 36 $LG_p^\ell$ modes, generated by combinations of $\ell\in[0,5]$ and $p\in[0,5]$. The images were recorded with a Charge-Couple Device (CCD) camera (6.5 $\mu$m pixel size) at the plane $z=0$, the plane of the ground glass, where we corroborated their spot size experimentally. As can be seen, their spot size increases with $p$ and $\ell$, as predicted by Eq. \ref{LGwaist}. In order to confirm that indeed, the structure of the beam does not influence the speckle size, we also encoded on the SLM a set of 36 apertures with homogeneous intensity and radius $r=\omega_0\sqrt{2p+|\ell|+1}$. Figure \ref{LGIntensity}(b) shows their recorded intensities at the plane $z=0$, for comparison with the LG modes they have been labeled using the modal indices $p$ and $\ell$ to indicate that their spot size is identical.

The speckle intensity generated by the ground glass was then recorded at a distance L from the scattering surface using the same CCD camera. Figure \ref{LGIntensity}(c) show the speckle obtained $LG_p^\ell$ modes, which clearly shows a decrease in size as $\ell$ increases while keeping $p$ constant (see Fig. \ref{LGIntensity}(c), rows). A similar behaviour is observed for increasing values of $p$ while keeping $\ell$ constant (see Fig. \ref{LGIntensity}(c), columns). For comparison, Fig. \ref{LGIntensity}(d) shows the speckle recorded when the homogeneous-intensity aperture illuminates the ground glass, again, the images are labeled using the modal numbers $p$ and$\ell$. Crucially, the speckle produced by both sets is almost identical, being impossible to discern which was generated with the apertures and which with the $LG_p^\ell$ modes. Nevertheless, a quantitative measure is required to compare the speckle size in both cases.

 \begin{figure}[bt]
\centering
\includegraphics[width=.45\textwidth]{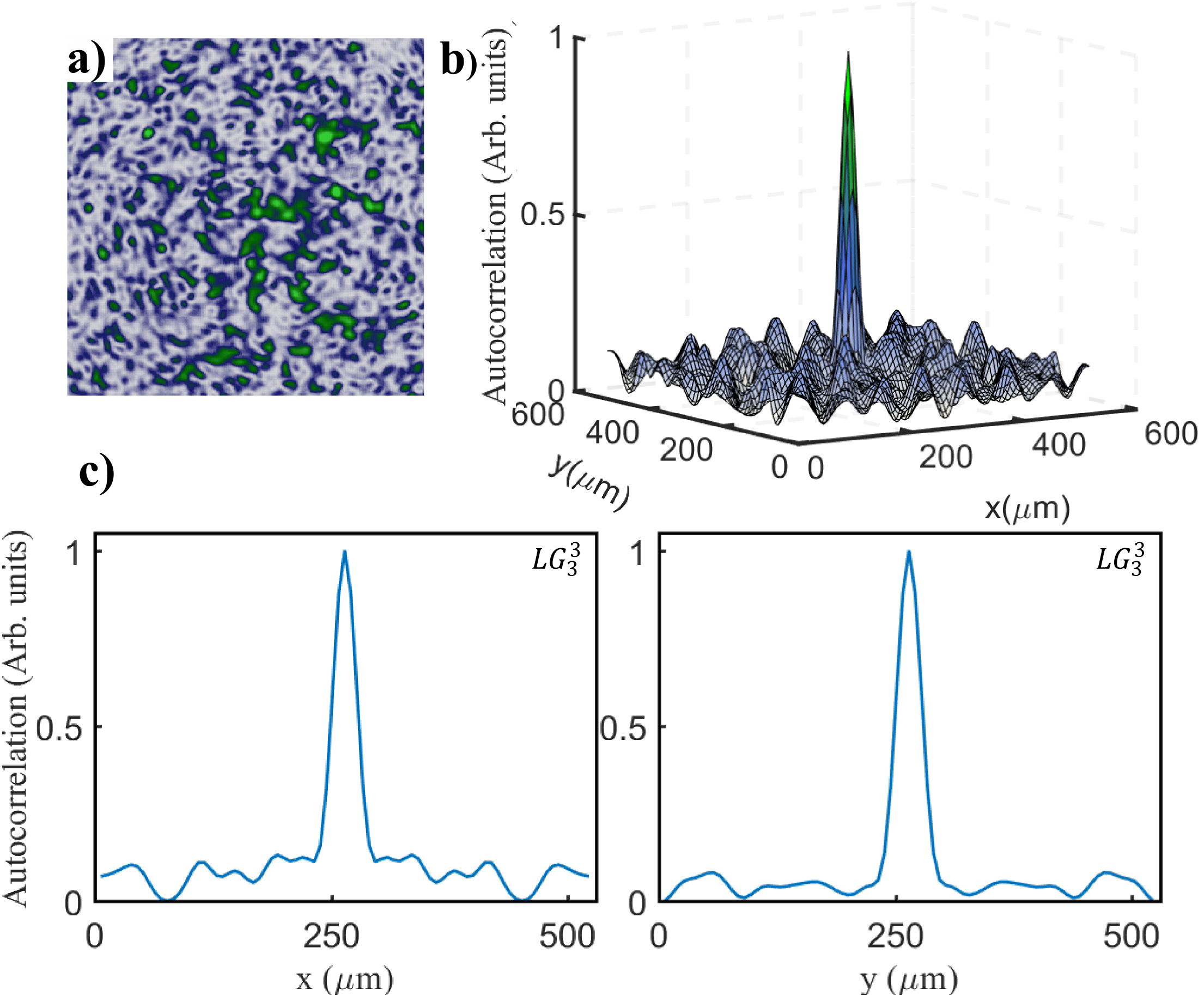}
\caption{(a) Example of an speckle pattern. (b) 3D autocorrelation of the speckle shown in (a). Transverse profile of the 3D autocorrelation shown in (b) along the $x$ (c) and $y$ (d) direction.}
\label{ImAutocorr}
\end{figure}

To measure the mean speckle size $\Delta s$, we performed a numerical autocorrelation of each speckle image with itself. Figure \ref{ImAutocorr}(a) shows an example of such a speckle image, while  Fig. \ref{ImAutocorr}(b) shows a 3D image of the autocorrelation, clearly showing a central peak, which correspond to the complete overlapping of speckle. Figures \ref{ImAutocorr}(c) and \ref{ImAutocorr}(d) show a transverse plot of the 3D autocorrelation image along the $x$ and $y$  axis, respectively. From each autocorrelation plot, we measure the speckle size as the full width at half-maximum (FWHM), and averaged along both directions. 
\begin{figure}[tb]
\centering
\includegraphics[width=.4\textwidth]{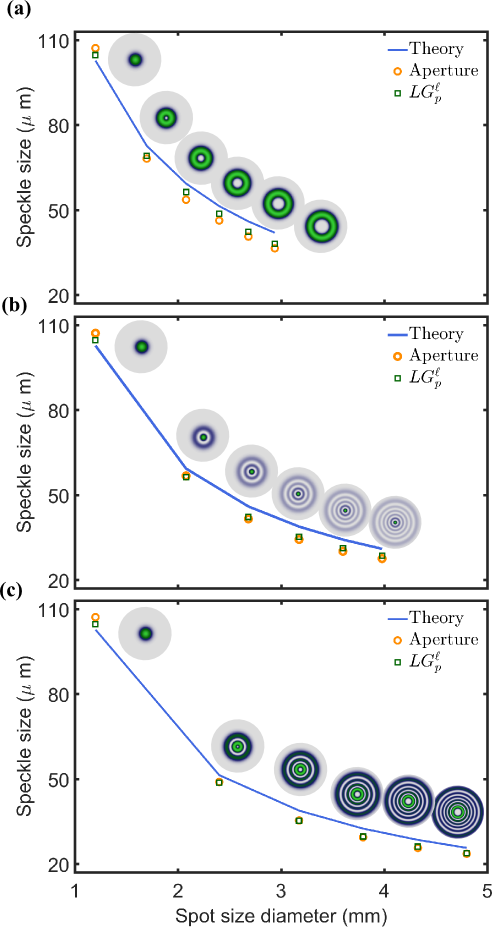}
\caption{Speckle size as function of the diameter of the illuminating spot size for both, the homogeneous-intensity aperture and the $LG_p^\ell$ modes. In (a) $p=0$, in (b) $\ell=0$ and in (c) $\ell\neq0$, $p\neq0$.}
\label{SpeckleSize}
\end{figure}
Our main results are presented in Fig. \ref{SpeckleSize} where we plot the speckle size as function of the mean spot-size diameter. Here we are comparing both cases, the aperture with homogeneous intensity and the $LG_p^\ell$ modes. Here, we only show three cases of interest: the case $p=0$ and $\ell\in[0,5]$ (Fig. \ref{SpeckleSize}(a)) the case $\ell=0$ and $p\in[0,5]$ (Fig. \ref{SpeckleSize}(b)) and the general case where both, $\ell$ and $p$ are non zero (Fig. \ref{SpeckleSize}(c)). In all cases, the solid line represents the speckle size as function of the spot size, as predicted from the theory given by Eq. \ref{SpSize} and \ref{LGss}, with $r=w_0\sqrt{2p+|\ell|+1}$. Notice the high resemblance between the theoretical curve, and the experimentally measured spot size for both the aperture and the $LG_p^\ell$ modes, proving that indeed the speckle size only depends on the illuminated area and not the shape of the same.

In conclusion, we analyzed the speckle produced by higher-order Laguerre-Gaussian modes, characterized by a non-homogeneous intensity distribution, and demonstrated that the mean speckle size is independent of the intricate  structure of the modes, as suggested by previous studies. We further provide with a mathematical expression that relates the mean speckle size with the modal number of the $LG_p^\ell$ modes. This expression was derived under the assumption that the illuminated area has a homogeneous intensity, we performed an experimental corroboration showing a good agreement between our mathematical expression and the experimentally measured speckle size. To further support our results, we performed additional experiments, where a circular aperture of homogeneous intensity was used as the illuminating source. Crucially, the speckle size generated with the $LG_p^\ell$ modes matches with very high accuracy the speckle size generated with the homogeneous-intensity aperture and to our proposed mathematical expression. These findings will be of great relevance in speckle-based applications as they path the way into the use of more complicated structured beams.

\section*{Funding Information}
 National Natural Science Foundation of China (NSFC) (11574065)

\begin{thebibliography}{10}
\newcommand{\enquote}[1]{``#1''}

\bibitem{Piederriere2005}
Y.~Piederri\`{e}re, F.~Boulvert, J.~Cariou, B.~L. Jeune, Y.~Guern, and G.~L.
  Brun, {\protect\JournalTitle{Opt. Express}} \textbf{13}, 5030 (2005).

\bibitem{Bianco2018}
V.~Bianco, P.~Memmolo, M.~Leo, S.~Montresor, C.~Distante, M.~Paturzo,
  P.~Picart, B.~Javidi, and P.~Ferraro, {\protect\JournalTitle{Light: Science
  \& Applications}} \textbf{7}, 48 (2018).

\bibitem{SanJuan2008}
J.~C. Ramirez-San-Juan, R.~Ramos-Garcia, I.~Guizar-Iturbide,
  G.~Martinez-Niconoff, and B.~Choi, {\protect\JournalTitle{Opt. Express}}
  \textbf{16}, 3197 (2008).

\bibitem{SanJuan2013}
J.~C. Ramirez-San-Juan, E.~Mendez-Aguilar, N.~Salazar-Hermenegildo,
  A.~Fuentes-Garcia, R.~Ramos-Garcia, and B.~Choi,
  {\protect\JournalTitle{Biomedical optics express}} \textbf{4}, 1883 (2013).

\bibitem{Francon1979}
M.~Francon, \emph{Laser Speckle and Applications in Optics} (Science Direct,
  1979).

\bibitem{Cheng2007}
H.~Cheng and T.~Q. Duong, {\protect\JournalTitle{Opt. Lett.}} \textbf{32}, 2188
  (2007).

\bibitem{Fercher1981}
A.~Fercher and J.~Briers, {\protect\JournalTitle{Optics Communications}}
  \textbf{37}, 326  (1981).

\bibitem{Heeman2019}
W.~Heeman, K.~Dijkstra, C.~Hoff, S.~Koopal, J.-P. Pierie, H.~Bouma, and E.~C.
  Boerma, {\protect\JournalTitle{Biomed. Opt. Express}} \textbf{10}, 2010
  (2019).

\bibitem{Zakharov2009}
P.~Zakharov, A.~V\"{o}lker, M.~Wyss, F.~Haiss, N.~Calcinaghi, C.~Zunzunegui,
  A.~Buck, F.~Scheffold, and B.~Weber, {\protect\JournalTitle{Opt. Express}}
  \textbf{17}, 13904 (2009).

\bibitem{Dunn2003}
A.~K. Dunn, A.~Devor, H.~Bolay, M.~L. Andermann, M.~A. Moskowitz, A.~M. Dale,
  and D.~A. Boas, {\protect\JournalTitle{Opt. Lett.}} \textbf{28}, 28 (2003).

\bibitem{Goldfischer1965}
L.~I. Goldfischer, {\protect\JournalTitle{J. Opt. Soc. Am.}} \textbf{55}, 247
  (1965).

\bibitem{Progresinoptics14}
J.~C. Dainty, \emph{Progress in optics}, vol. XIV (North Holland, 1976).

\bibitem{rosales2017shape}
C.~Rosales-Guzm{\'a}n and A.~Forbes, \emph{How to shape light with spatial
  light modulators} (SPIE, 2017).

\bibitem{Roadmap}
H.~Rubinsztein-Dunlop, A.~Forbes, M.~Berry, M.~Dennis, D.~L. Andrews,
  M.~Mansuripur, C.~Denz, C.~Alpmann, P.~Banzer, T.~Bauer \emph{et~al.},
  {\protect\JournalTitle{Journal of Optics}} \textbf{19}, 013001 (2016).

\bibitem{TwPh}
J.~P. Torres and L.~Torner, \emph{{Twisted Photons}} (Wiley-VCH, Bristol.,
  2011).

\bibitem{Hell1994}
S.~W. Hell and J.~Wichman, {\protect\JournalTitle{Optics Letters}} \textbf{19},
  780 (1994).

\bibitem{RosalesGuzman2013}
C.~Rosales-Guzm{\'a}n, N.~Hermosa, A.~Belmonte, and J.~P. Torres,
  {\protect\JournalTitle{Sci. Rep.}} \textbf{3}, 2815 (2013).

\bibitem{Trichili2016}
A.~Trichili, C.~Rosales-Guzm{\'{a}}n, A.~Dudley, B.~Ndagano, A.~{Ben Salem},
  M.~Zghal, and A.~Forbes, {\protect\JournalTitle{Scientific Reports}}
  \textbf{6}, 27674 (2016).

\bibitem{Willner2015}
A.~E. Willner, H.~Huang, Y.~Yan, Y.~Ren, N.~Ahmed, G.~Xie, C.~Bao, L.~Li,
  Y.~Cao, Z.~Zhao, J.~Wang, M.~P.~J. Lavery, M.~Tur, S.~Ramachandran, A.~F.
  Molisch, N.~Ashrafi, and S.~Ashrafi, {\protect\JournalTitle{Advances in
  Optics and Photonics}} \textbf{7}, 66 (2015).

\bibitem{Cox2016}
M.~A. Cox, C.~Rosales-Guzm\'{a}n, M.~P.~J. Lavery, D.~J. Versfeld, and
  A.~Forbes, {\protect\JournalTitle{Optics Express}} \textbf{24}, 18105 (2016).

\bibitem{Malik2012}
M.~Malik, M.~O'Sullivan, B.~Rodenburg, M.~Mirhosseini, J.~Leach, M.~P.~J.
  Lavery, M.~J. Padgett, and R.~W. Boyd, {\protect\JournalTitle{Opt. Express}}
  \textbf{20}, 13195 (2012).

\bibitem{Cox2018}
M.~A. Cox, L.~Cheng, C.~Rosales-Guzm\'an, and A.~Forbes,
  {\protect\JournalTitle{Phys. Rev. Applied}} \textbf{10}, 024020 (2018).

\bibitem{Allen92}
L.~Allen, M.~W. Beijersbergen, R.~J.~C. Spreeuw, and J.~P. Woerdman,
  {\protect\JournalTitle{Physical Review A}} \textbf{45}, 8185 (1992).

\bibitem{Reddy2014}
S.~G. Reddy, S.~Prabhakar, A.~Kumar, J.~Banerji, and R.~P. Singh,
  {\protect\JournalTitle{Opt. Lett.}} \textbf{39}, 4364 (2014).

\bibitem{Reddy2016}
S.~G. Reddy, C.~P, P.~Vaity, A.~Aadhi, S.~Prabhakar, and R.~P. Singh,
  {\protect\JournalTitle{Journal of Optics}} \textbf{18}, 055602 (2016).

\bibitem{RosalesGuman2017a}
C.~Rosales-Guzm{\'{a}}n, N.~Bhebhe, and A.~Forbes,
  {\protect\JournalTitle{Journal of Optics}} \textbf{25}, 25697 (2017).

\bibitem{Goodman1975}
J.~W. Goodman, \emph{Laser Speckle and Related Phenomena}, vol.~9
  (Springer-Verlag Berlin Heidelberg, 1975).

\bibitem{MatlabTutorial}
D.~G. Voelz, \emph{Computational Fourier Optics: A MATLAB Tutorial}, vol. TT89
  of \emph{Tutorial Texts (Book 89)} (SPIE Press, 2011).

\bibitem{Phillips1983}
R.~L. Phillips and L.~C. Andrews, {\protect\JournalTitle{Appl. Opt.}}
  \textbf{22}, 643 (1983).

\end{thebibliography}

\end{document}